\documentclass[12pt,preprint]{aastex}

\newcommand{\europhysl}{Europhys. Lett.}
\newcommand{\nphysb}{Nuclear Phys. B}
\newcommand{\cmp}{Commun. Math. Phys.}

\shorttitle{}
\shortauthors{Trott \& Melatos}

\begin{document}


\title{Collapsed and Extended Cold Dark Matter Haloes in Softened $N$-Body Gravity}


\author{C. M. Trott\altaffilmark{1} and A. Melatos\altaffilmark{1}}
\affil{School of Physics, University of Melbourne, Victoria 3010, Australia}


\altaffiltext{1}{Email: ctrott@physics.unimelb.edu.au, a.melatos@physics.unimelb.edu.au}

\begin{abstract}
The statistical mechanics of $N$ cold dark matter (CDM) particles interacting via a softened gravitational potential is reviewed in the microcanonical ensemble and mean-field limit. A phase diagram for the system is computed as a function of the total energy $E$ and gravitational softening length $\epsilon$. For softened systems, two stable phases exist: a collapsed phase, whose radial density profile $\rho(r)$ is a central Dirac cusp, and an extended phase, for which $\rho(r)$ has a central core and $\rho(r)$ $\sim$ $r^{-2.2}$ at large $r$. It is shown that many $N$-body simulations of CDM haloes in the literature inadvertently sample the collapsed phase only, even though this phase is unstable when there is zero softening. Consequently, there is no immediate reason to expect agreement between simulated and observed profiles unless the gravitational potential is appreciably softened in nature.
\end{abstract}

\keywords{dark matter $-$ galaxies: kinematics and dynamics $-$ galaxies: structure $-$ gravitation}

\section{Introduction}

Cold dark matter (CDM) theory successfully describes many aspects of the formation of large-scale structure in the universe \citep{peebles82,davis85}. However, mismatches do exist between its predictions and observations, such as the cusp-core controversy, missing satellites \citep{klypin99,moore99a} and the angular momentum problem \citep{nav91,thacker01}. In particular, the cusp-core issue has provoked much debate. 
CDM simulations consistently yield density profiles with steeper inner slopes  (power-law exponent between $-$1 and $-$1.5) than observational studies which have found a range of slopes, including constant density cores in dark matter dominated low surface brightness galaxies \citep{deblok01} and shallow slopes in clusters with gravitationally lensed arcs  \citep{sand03}. 
These results, among others, have initiated discussion about the role baryons play in softening simulated cores \citep{athan,shen03} and the observational effects that may mask cusps in low surface brightness galaxies \citep{deblok01,swaters03}. Recent $N$-body results demonstrate that, at the current resolution of simulations, the central power-law exponent does not converge to a universal value \citep{nav03}.

In numerical simulations, a softened gravitational potential is used to prevent the macro-particles (10$^5$--10$^7M_\odot$) from experiencing artificially strong two-body interactions \citep[for example]{nfw96}. The softening length, $\epsilon$, is chosen to maximise the resolution while suppressing two-body effects over the simulation running time. In view of the ongoing disagreement regarding the form of $\rho(r)$, it is important to clarify \textit{analytically}, by a code-independent argument, whether the choice of $\epsilon$ affects the physics of the system and hence $\rho(r)$. 
In this paper, we employ the framework of statistical mechanics \citep{pad90}, drawing upon recent results on phase transitions in $N$-body systems with attractive power-law potentials \citep{ispolcohen01,vegasan02}. Self-gravitating particles behave qualitatively differently to many other statistical systems because gravity is an unscreened, long-range force. They are best examined within the microcanonical ensemble, where the energy and number of particles are fixed and phases with negative specific heat are allowed. We apply the results of the classical theory of self-gravitating, $N$-body systems to demonstrate the effects of introducing a short distance cutoff in numerical simulations, in particular the effect on stability.

The study of the thermal stability of self-gravitating systems has a long history. 
\citet{lynden68} showed that spherical systems of point particles in a box with reflecting walls are gravitationally unstable below a critical temperature, collapsing catastrophically to a central point.
\citet{aronson72} generalised this work to a spherical system of $N$ classical hard spheres in contact with a heat bath, showing the gravothermal instability to be a general feature of self-gravitating systems held at a constant temperature.
\citet{hertel71}, investigating point fermions obeying the Pauli Exclusion Principle, showed that a {\em stable} low temperature phase can exist if the gravitational potential is softened, transforming the gravothermal instability to a phase transition to the low temperature phase. 
In this paper, we extend these results and use them to reinterpret some of the ambiguous results of numerical 
simulations of CDM haloes discussed above.
A detailed comparison with preceding analytic work is presented in Section \ref{comparison}.

Section 2 briefly reviews the formalism for treating $N$ self-gravitating, collisionless particles statistically. In Section 3, we apply the formalism to compute $\rho(r)$ analytically as a function of $E$, the total energy, and $\epsilon$. The result is a thermodynamic phase diagram that contains both collapsed and extended haloes. In Section 4, we locate published $N$-body simulations on the phase diagram and show they are biased towards the collapsed phase. This phase is unstable for $\epsilon=0$, suggesting that collapsed haloes are an artificial by-product of the softened potential; there is no immediate reason to expect agreement between simulated and observed profiles unless the gravitational potential is appreciably softened in nature. We emphasise at the outset that it is not our intention to reproduce realistic CDM haloes with non-zero angular momentum and hierarchical clustering; rather, we demonstrate in a code-independent manner how the softening used in $N$-body simulations may artificially alter the density profiles found.

\section{Statistical mechanics of $N$ self-gravitating particles}

\subsection{Density of states in the microcanonical ensemble}

The properties (e.g. energy, entropy) and collective behaviour (e.g. gravothermal catastrophe) of a self-gravitating gas of CDM particles in thermodynamic equilibrium take different values when computed in different statistical ensembles because the long-range nature of the gravitational potential renders the system inseparable from its environment \citep{pad90,ispolcohen01,vegasan02}. In this paper, we follow previous studies by considering the self-gravitating gas in the microcanonical ensemble (MCE), whose features are constant energy, volume and particle number. Particles do not evaporate from the system over time and the walls of the container are perfectly reflecting. 
The MCE is more appropriate than the canonical ensemble (CE) for three reasons: (i) it is unclear how to construct an external heat bath (required by the CE) for a long-range potential, because the system interferes with the environment \citep{huang87}; (ii) states with negative specific heat are inaccessible in the CE \citep{pad90}; and (iii) the equilibrium density profile in the violently relaxed (Smoluchowski) limit is the singular isothermal sphere in the CE, contrary to observations \citep{sire02}.

The density of states, $g(E)$, is the volume of the (6$N-$1)-dimensional surface of constant energy $E$ in phase space (${\bf{x}}_1,...,{\bf{x}}_N,{\bf{p}}_1,...,{\bf{p}}_N$), where (${\bf{x}}_i,{\bf{p}}_i$) are the co-ordinates and momenta of the $i$-th particle. 
At any one moment, the system occupies one point in the 6$N$-dimensional phase space. 
For particles of equal mass $m$, one has

\begin{equation}\label{g}
g(E) = \frac{1}{N!}\int\delta\left[E-\sum_{i=1}^{N}\frac{p_i^2}{2m}-\sum_{i{\neq}j}^{N}V({\bf{x}}_i,{\bf{x}}_j)\right]d^{3N}pd^{3N}x,
\end{equation}
where the first and second sums give the kinetic and potential energy, and the integral is over phase space volume. The gravitational potential, $V$, is given by $V=-Gm^2|{\bf{x}}_i-{\bf{x}}_j|^{-1}$ or, if the potential is artificially softened over a characteristic length $\zeta$, by $V=-Gm^2[({\bf{x}}_i-{\bf{x}}_j)^2+\zeta^2]^{-1/2}$.

The thermodynamic entropy $S$ (up to a constant) and the temperature $T$ of the system are defined in terms of $g(E)$:

\begin{equation}
S(E) = k_B \ln{g(E)},
\end{equation}

\begin{equation}
\beta(E) = \frac{1}{k_BT} = \frac{\partial{S(E)}}{\partial{E}}.
\end{equation}
These quantities are hard to interpret when assigned to a system far from equilibrium. Note that $g(E)$ diverges for $\zeta=0$ and $N>2$; any two particles can be brought arbitrarily close together, liberating an infinite amount of potential energy, so that the co-ordinate space integral diverges \citep{pad90}. This is a serious problem because it is impossible to achieve thermodynamic equilibrium if $g(E)$ diverges; the system does not have time to sample the infinite number of possible microstates with equal probability \citep{chabanol00}. 
If the dark matter particles are fermions, the Pauli Exclusion Principle does prevent this problem. However, the fraction of the phase space volume sampled by $N$ mildly relativistic CDM particles in a time $t$, given by $(ct/R)^{3N}({\zeta}c^2/2GmN^{4/3})^{3N/2}$, is exceedingly small for most proposed CDM particles, e.g. $\zeta^2/m \sim 8{\times}10^{-(29-26)}$m$^2$/GeV for self-interacting dark matter \citep{spergel00}.

% \textit{Derivation of the $g(E)$ for the MCE and the mean-field theorem (maybe put this in the appendix) and an explanation of the limitations of the approximation. de Vega \& Siebert (2002) discuss this and use monte carlo simulations to look at the system away from the mean-field approximation.}

\subsection{Integral equations for $\rho(r)$ in the mean-field limit} \label{meanfield}

The density of states is evaluated in the continuum (mean-field) limit by integrating over momentum and then expressing  the remaining configurations as a functional integral over possible density profiles $\rho(x)$ \citep{vegasan02,ispolcohen01},

\begin{equation}\label{functional}
g(E) = {\int}D\rho\int^{+\infty}_{-\infty}\frac{d\gamma}{2{\pi}i}\int^{+{\infty}}_{-{\infty}}\frac{d\beta}{2{\pi}i}\exp[Ns(\rho,\xi,\gamma,\beta)],
\end{equation}
where the effective dimensionless action

\begin{eqnarray}\label{action}
s(\rho,\xi,\gamma,\beta) &=& \beta\xi + \frac{\beta}{2}\int\int\frac{\rho({\bf{x}}_1)\rho({\bf{x}}_2)}{|{\bf{x}}_1-{\bf{x}}_2|^2}d^3{\bf{x}}_1d^3{\bf{x}}_2\\\nonumber
&+& \gamma\int\rho({\bf{x}})d^3{\bf{x}}-\gamma-\frac{3}{2}\ln\beta-\int\rho({\bf{x}})\ln\rho({\bf{x}})d^3{\bf{x}}
\end{eqnarray}
and dimensionless energy

\begin{equation}
\xi = ER/GM^2
\end{equation}
are quantities defined by \citet{ispolcohen01}. In (\ref{functional}) and (\ref{action}), and throughout the remainder of this paper, the density profile and position co-ordinates ${\bf{x}}$ are written as dimensionless quantities, relative to the total mass $M=Nm$ and outer radius $R$ of the system, with $x={|\bf{x}|}/R$ and $\epsilon=\zeta/R$.

Upon evaluating the functional integral by a saddle point method (which involves extremising the action), (\ref{functional}) reduces to three coupled integral equations describing the density profile $\rho(x)$, the central density $\rho_0$ and the inverse temperature $\beta = 1/k_BT$. For a Newtonian potential, one has

%\begin{mathletters}\label{coupled}
\begin{eqnarray}\label{coupled}
&&\rho(x)=\rho_0 \exp\left[\frac{2\pi\beta}{x}\int^{1}_{0}\rho(x_1)x_1(|x+x_1|-|x-x_1|)dx_1\right], \\\label{coupled2}
&&\frac{1}{\rho_0}=\int^{1}_{0} 4{\pi}{x^2_2} \exp\left[\frac{2\pi\beta}{x_2}\int^{1}_{0}\rho(x_1)x_1(|x_2+x_1|-|x_2-x_1|){dx_1}\right] dx_2, \\\label{coupled3}
&&\frac{3}{2\beta}=\xi + 4\pi^2\int^{1}_{0}\int^{1}_{0}\rho(x_1)\rho(x_2)x_1x_2(|x_1+x_2|-|x_1-x_2|)dx_1dx_2.
\end{eqnarray}
%\end{mathletters}

\noindent Note that the factor $4{\pi}x^2_2$ in (\ref{coupled2}) was omitted due to a typographical error by \citet{ispolcohen01}. The solutions to these equations describe the density profile, entropy and temperature of an equilibrium system for a given energy. To obtain analogous equations for the softened potential, we replace $|...|$ everywhere with $[(...)^2+\epsilon^2]^{1/2}$. This extension is valid in the mean-field formalism for $\epsilon$ small: correction terms are $O(\epsilon)$ \citep{vegasan02}.

The mean-field limit is only meaningful physically for $\epsilon{\neq}0$, otherwise $g(E)$ diverges. \citet{vegasan02} verified the mean-field results against Monte-Carlo simulations and an alternative analytic method known as the Mayer cluster expansion, where the density of states is expanded as a combinatorial series in the dilute limit ($\xi\gg{1}$). The different approaches are in accord in the dilute limit. In order to verify the mean-field approach in the high-density limit ($\xi\ll{1}$), relevant to the extended and collapsed phases studied here, we need to calculate the correction terms in this limit, following \citet{vegasan02} --- a project outside the scope of this paper. 
Nevertheless, to give a rough idea of these corrections, we note (by analogy) that they are of order $O(\eta^2\epsilon, \eta\epsilon^2)$ in the CE, where $\eta=Gm^2N/RT$ in the CE is a proxy for $\xi^{-1}$ in the MCE. Note that the classical thermodynamic limit ($N/V$ constant as $N,V\rightarrow\infty$) does not apply for gravitating systems; thermodynamic quantities are finite if proportional to $N/V^{1/3}$ as $N,V\rightarrow\infty$.

\section{Radial density profile of a CDM halo}\label{profile}

We solve (\ref{coupled})--(\ref{coupled3}) for the radial density profile $\rho(x)$ by the following iterative relaxation scheme \citep{ispolcohen01}: given the current iterate of the profile, $\rho^{(i)}(x)$, apply (\ref{coupled3}), (\ref{coupled2}) and (\ref{coupled}) to compute $\beta^{(i+1)}$, $\rho_0^{(i+1)}$ and $\rho^*(x)$ in that order, then apply $\rho^{(i+1)}(x) = {\sigma}\rho^*(x) + (1-\sigma)\rho^{(i)}(x)$ until the convergence criterion

\begin{equation}
4\pi\int^1_0 |\rho^{(i+1)}(x)-\rho^{(i)}(x)|x^2dx < \delta,
\end{equation}
is satisfied. We typically adopt $\delta=10^{-6}$ and $0.01\leq\sigma\leq1$ ($\delta \ll \sigma$) in this work. The softening can be introduced into this scheme in two ways: (i) as a nonzero lower limit of integration in the integrals in (\ref{coupled})--(\ref{coupled3}); and (ii) in the potential, $V=-Gm^2[({\bf{x}}_i-{\bf{x}}_j)^2+\epsilon^2]^{-1/2}$. Both approaches were tested and found to produce qualitatively similar behaviour; we concentrate on the latter in this work as it is more closely allied to $N$-body simulations.

\subsection{Stable versus unstable phases: $\epsilon=0$}

\begin{figure}
\plottwo{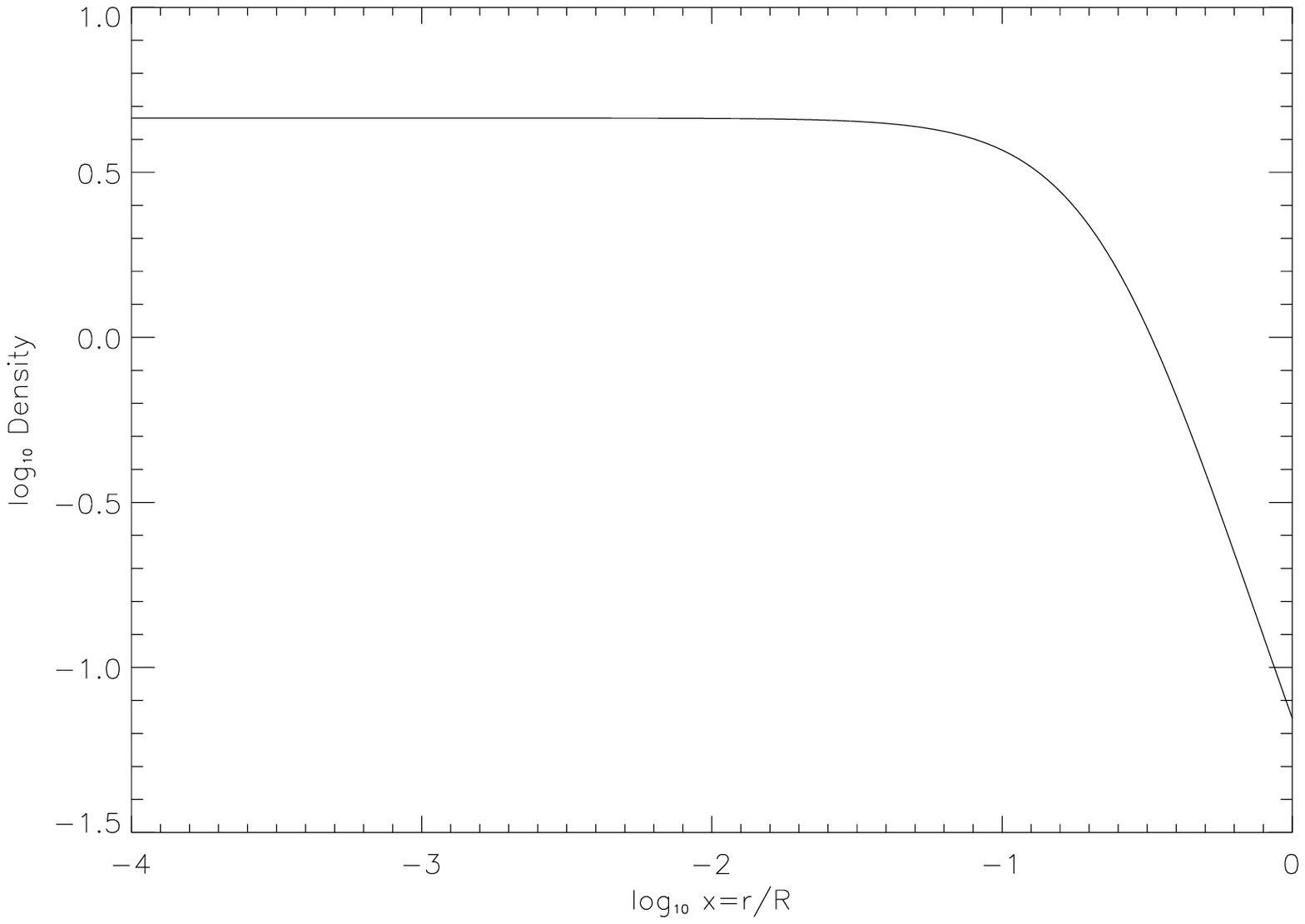}{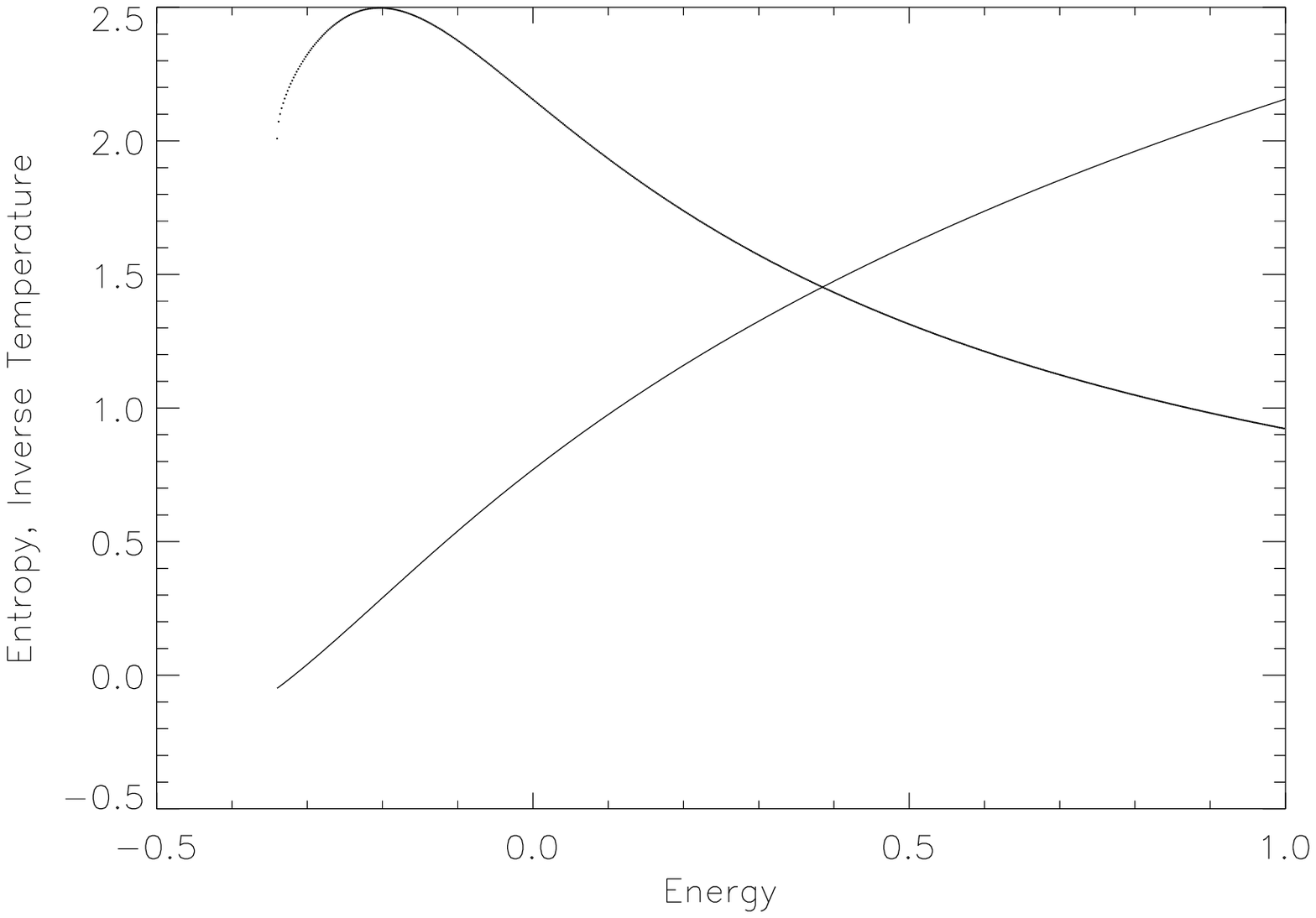}
\caption{(a) Equilibrium radial density profile for a system with $\xi > \xi_c$ and $\epsilon=0$. The profile is similar to a softened isothermal sphere, but with $\rho(x)\propto{x^{-2.2}}$ as $x\rightarrow{1}$. (b) Dimensionless entropy (monotonically increasing) and inverse temperature (peaked curve) as a function of energy for $\xi > \xi_c$ when the gravitational potential is not softened. Below $\xi_c{\simeq}-0.335$, no stable solution exists. \label{nice_density}}
\end{figure}

With no softening present in the gravitational potential, a stable solution of (\ref{coupled})--(\ref{coupled3}) formally exists above a cutoff energy $\xi>\xi_c\simeq-0.335$. The density profile of the halo exhibits a flat central core, with $d\rho/dx\rightarrow0$ as $x\rightarrow0$, and near-isothermal wings, with $\rho(x)$ $\propto$ $x^{-\alpha}$ ($\alpha\simeq$2.2) as $x\rightarrow\infty$, as illustrated in Figure \ref{nice_density}(a). 
We refer to it as the {\em extended} phase.
It agrees with the solution for secondary infall onto a spherical perturbation \citep{bert85} and behaves asymptotically like the spherical, thermally conducting polytrope \citep{lyndeneggle80} and infinite-dimensional Brownian gas \citep{sire02}.

For $\xi<\xi_c$ and $\epsilon=0$, a formal solution of (\ref{coupled})--(\ref{coupled3}) does not exist. 
The entropy and temperature are discontinuous at this cutoff energy as shown in Figure \ref{nice_density}(b). This is the well-known gravothermal catastrophe \citep{ant62}. Note that, for $\xi=-1/4$, the singular isothermal sphere $\rho(x)=(4{\pi}x^2)^{-1}$, $\rho_0=(4{\pi}e^2)^{-1}$ and $\beta=2$ is always a solution of (\ref{coupled})--(\ref{coupled3}), as can be verified analytically, but it is not stable and so the iterative procedure never converges to it, but rather to Figure \ref{nice_density}(a).

%\clearpage

%\clearpage

\subsection{Collapsed versus extended phases: $\epsilon\neq0$}\label{comparison}

\begin{figure}
\plottwo{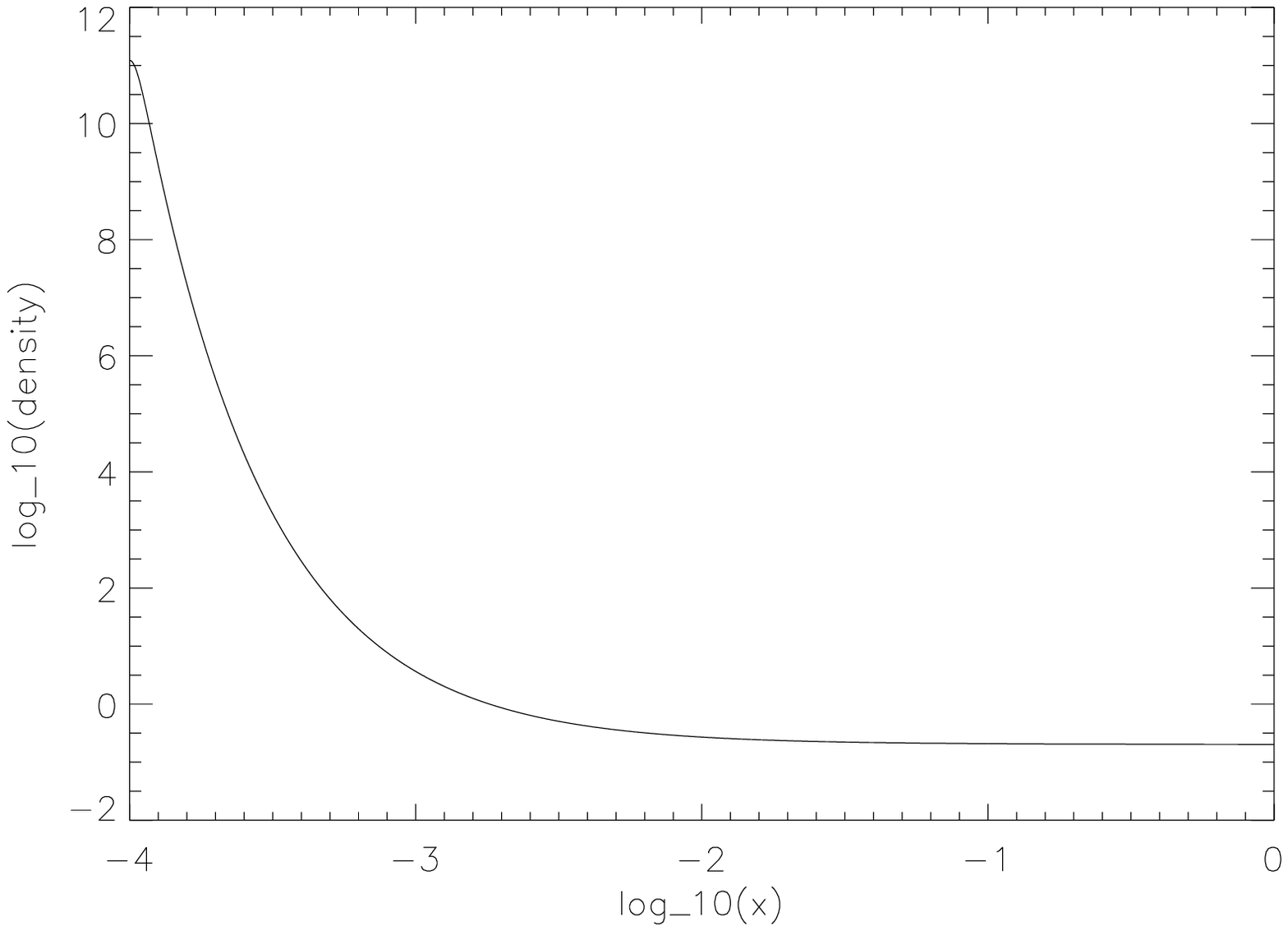}{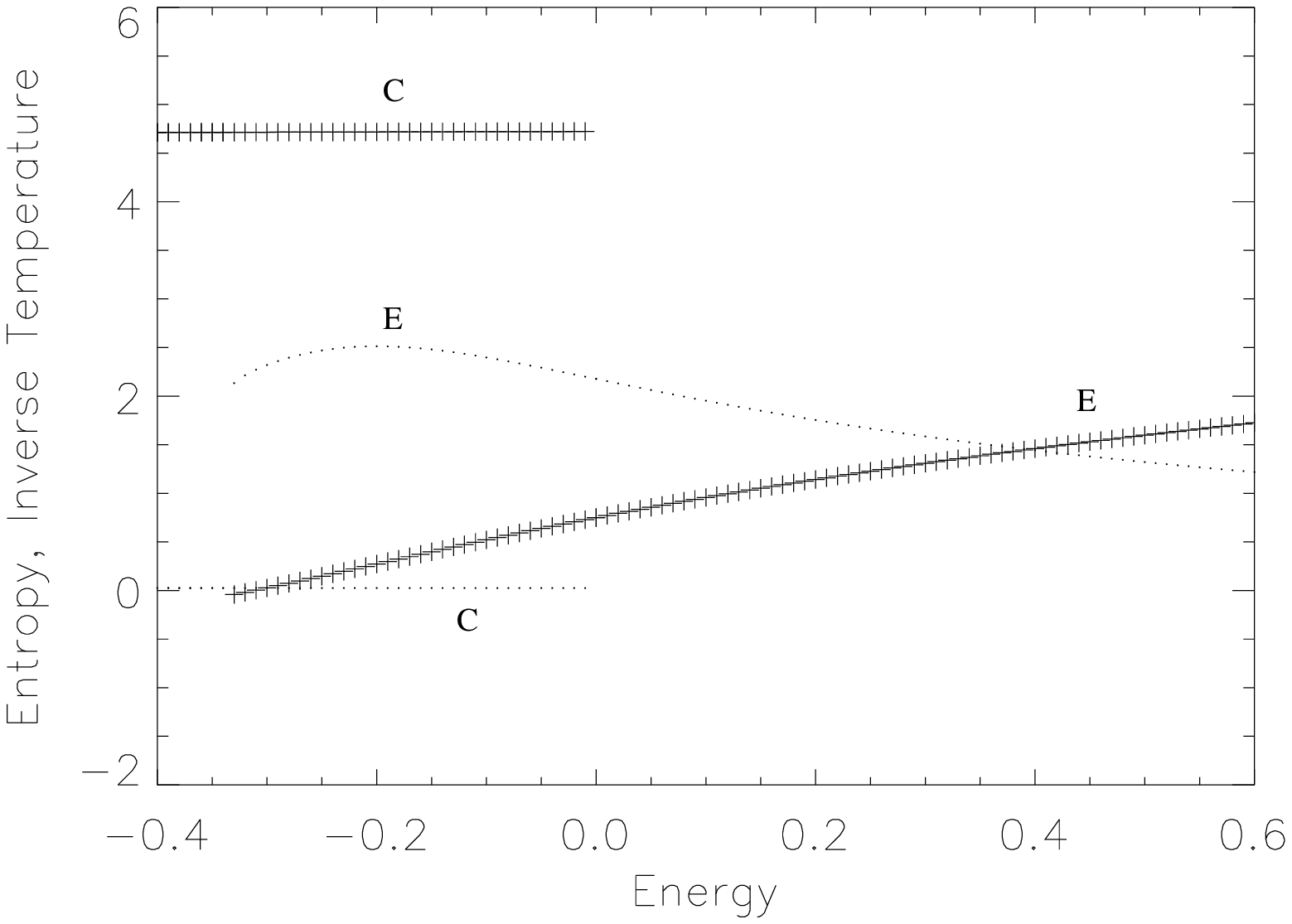}
\caption{(a) Radial density profile for the collapsed phase of a potential with softening $\epsilon=10^{-4}$. (b) Dimensionless entropy (thick curve) and inverse temperature (thin) versus energy. The collapsed and extended phases are labelled C and E respectively. Both quantities display discontinuous jumps at the phase transitions at $\xi_c\simeq-0.335$ and $\xi_c^{(+)}\simeq0.0$. If the energy is increased from below $\xi_c$, the system remains in the collapsed phase until $\xi=\xi_c^{(+)}$, when it jumps to the extended phase. If the energy is reduced from above $\xi_c^{(+)}$, the system remains in the extended phase until $\xi=\xi_c$, when it jumps to the collapsed phase. \label{collapse_density}}
\end{figure}

If the gravitational potential is softened, a stable phase exists for all values of $\xi$. For $\xi>\xi_c^{(+)}\simeq0$, the halo is extended as in Figure \ref{nice_density}(a), with a flat core and near-isothermal envelope, $\rho(x)\propto{x}^{-2.2}$. However, for $\xi<\xi_c$, the halo is collapsed. Figure \ref{collapse_density}(a) displays the density profile of such a collapsed halo for $\epsilon=10^{-4}$ (note that both axes are logarithmic). The halo has a steep Dirac peak (`cusp') at $x=0$: $\rho(x)$ is flat for $x\lesssim\epsilon$, decreases as a large inverse power of $x$ for $x\gtrsim\epsilon$ and flattens for $x\rightarrow{1}$. For intermediate energies in the range $\xi_c\leq\xi\leq\xi_c^{(+)}$, the system is bistable: the halo can be either extended or collapsed depending on the initial conditions and the route to equilibrium. 
Figure \ref{collapse_density}(b), a plot of entropy and temperature as a function of energy, illustrates this bistability and the hysteresis to which it can lead. $S(\xi)$ and $\beta(\xi)$ jump discontinuously at both $\xi_c^{(+)}$ and $\xi_c$. If $\xi$ enters the intermediate range from below, the halo remains collapsed until $\xi$ exceeds $\xi_c^{(+)}$. Alternatively, if $\xi$ enters from above, the halo remains extended until $\xi$ is reduced below $\xi_c$ \citep{ispolcohen01}.

The critical energy $\xi_c$, and the collapsed and extended profiles we obtain,
are consistent with previous analyses \citep{ant62,aronson72,pad90}.
For example, \citet{aronson72} find a phase transition at $\xi\simeq{-0.3}$, consistent with our value $\xi_c\simeq{-0.335}$, and a critical reciprocal temperature at the transition in the range $\beta_c = 1.2-1.6$ for $\epsilon = 10^{-20}$, consistent with $\beta\sim2$ in this paper ($\epsilon=10^{-4}$). 
A more precise comparison is prohibited by the adoption of the CE rather than the MCE
in most previous work.
The van der Waals model proposed by \citet{pad90} is an exception;
it is examined in detail in the following section.
The phenomenon of bistability was overlooked until the work of \citet{ispolcohen01}.

%\clearpage

%\clearpage

The behaviour of the system depends somewhat on the choice of the relaxation parameter $\sigma$, defined at the start of this section. Table \ref{phase_table} displays the minimum softening length for which the system makes the transition to a stable collapsed phase, for a given value of $\sigma$. The phase transition from the extended to the collapsed phase is increasingly delicate as $\epsilon$ decreases: for larger values of $\sigma$, the system is less likely to reach the critical point where the phase transition occurs and thus remains in the extended phase. Although the effect is numerical, it potentially reflects the relative likelihoods of the possible routes that the real system can take to equilibrium.

%\clearpage

\begin{table}
\centering
\begin{tabular}{ll}\hline
{$\sigma$} & {$\epsilon_{\rm{min}}$} \\\hline
{0.01} & {$10^{-14}$}\\
{0.03} & {$10^{-10}$}\\
{0.05} & {$10^{-6}$}\\
{0.1} & {$10^{-2}$}\\
\end{tabular}
\caption{Minimum softening, $\epsilon_{\rm{min}}$, for which a stable collapsed solution is found for a given relaxation parameter $\sigma$. \label{phase_table}}
\end{table}

%\clearpage

\subsection{van der Waals equation of state}
An alternative, phenomenological way to model a CDM halo 
interacting via a softened gravitational potential 
is to solve the Lane-Emden equation for a nonideal,
isothermal gas obeying a van der Waals 
equation of state,
$P\propto \rho T (1-\rho/\rho_{\rm m})^{-1}$,
where $P$ denotes the pressure and $\rho_{\rm m}$ is the maximum density
allowed by hard-sphere packing of the CDM particles
\citep{aronson72,pad90}.
The analogy with a softened gravitational potential implies
$\rho_{\rm m} \sim m \epsilon^{-3}$,
although it is clear from the outset that this analogy is inexact;
the hard-core, van der Waals potential effectively excludes a volume
$\sim \epsilon^3$ around each particle,
whereas softened gravity suppresses the mutual acceleration of particles
separated by a distance $\epsilon$ without preventing them
from `coasting' even closer together.
We compare our solutions of (\ref{coupled})--(\ref{coupled3}) 
with the van der Waals model
by integrating the nonideal Lane-Emden equation
\begin{equation}
 \frac{1}{r^2} \frac{d}{dr}
 \left[ \frac{r^2}{\rho} \frac{d}{dr} 
  \left(
   \frac{\rho T}{1 - \rho/\rho_{\rm m}}
  \right)
 \right]
 = - 4\pi G \rho
\label{eq:laneemden}
\end{equation}
from $r=R_1$ to $r=R$, obtaining
\begin{equation}
 \frac{1}{\rho(R)} 
  \left. \frac{d\rho}{dr} \right|_R
 - \frac{R_1^2} {R^2 \rho(R_1)
  [ 1 - \rho(R_1)/\rho_{\rm m} ]^{2} }
  \left. \frac{d\rho}{dr} \right|_{R_1}
 =
 - \beta M_1 / M~,
\label{eq:intlaneemden}
\end{equation}
where $M_1$ denotes the mass enclosed in the volume $R_1\leq r \leq R$.
[We use $\rho(R)\ll\rho_{\rm m}$ to simplify (\ref{eq:intlaneemden}).]
The quantities $\rho(R)$, $\rho(R_1)$, their derivatives,
$M_1$, and $\beta$ can be extracted from our numerically computed
profiles, which satisfy (\ref{coupled})--(\ref{coupled3}),
in order to find $\rho_{\rm m}$ as a function of $\xi$ and $\epsilon$
and hence compare with the results of \citet{aronson72} and 
\citet{pad90}.
The inner integration limit $R_1$ must be positioned carefully
near the inflection point of the central Dirac peak,
e.g.\ at $r = 10^{-3.8}$ in Figure \ref{collapse_density},
in order to avoid the innermost grid cells,
where the numerical solution is noisiest,
while ensuring that $\rho(R_1)/\rho_{\rm m}$ is not too small,
to avoid roundoff error when solving
(\ref{eq:intlaneemden}) for $\rho_{\rm m}$.

\begin{figure}
\plotone{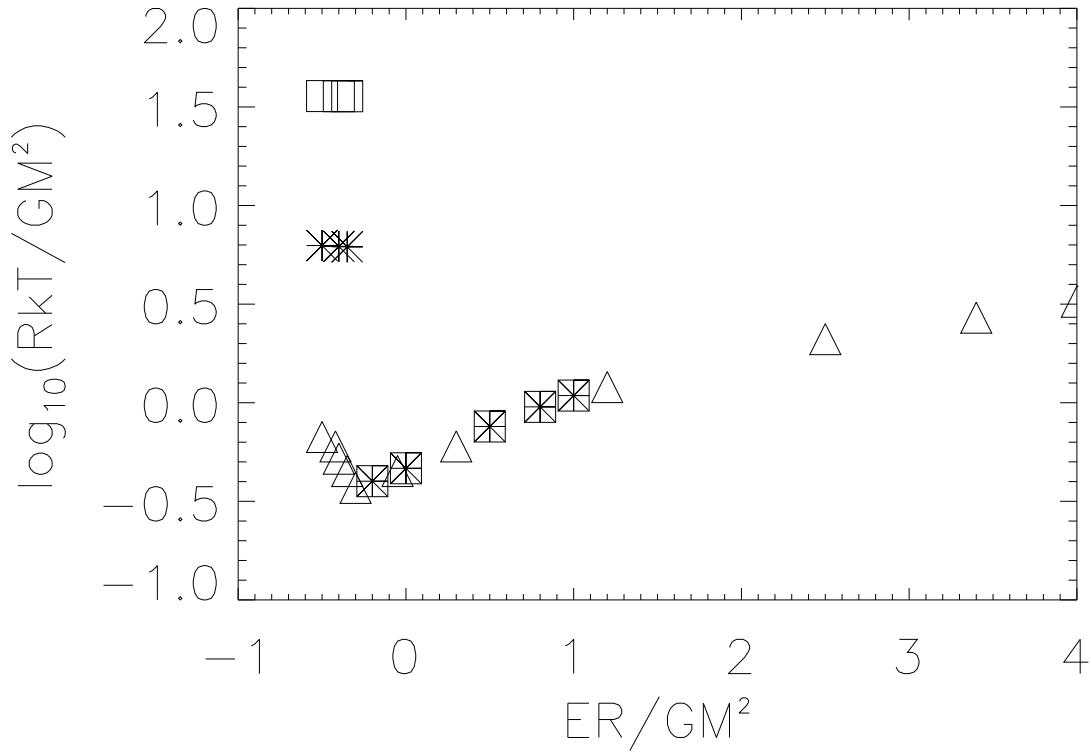}
\caption{Temperature $T$ as a function of total mechanical energy $E$
for a range of softening lengths $\epsilon$.
The open squares and asterisks denote CDM haloes with
$\epsilon=10^{-4}$ and $10^{-3}$ respectively,
modeled by the density of states formalism (\ref{coupled})--(\ref{coupled3}).
These haloes
correspond to $\log a = -10.1$ and $-9.00$ in the van der Waals formalism.
The open triangles denote CDM haloes with $-4\leq \log a \leq -2$,
modeled by the van der Waals formalism \citep{pad90}. 
For any given $E$ in the plotted range, 
$T$ is the same for all values $-4\leq \log a \leq -2$;
moreover, these $a$ values are systematically greater than the
predictions of the density of states formalism.
\label{vanderwaals}}
\end{figure}

In Figure \ref{vanderwaals}, we compare the
density of states and van der Waals models
by plotting $1/\beta$ versus $\xi$ for both models
in the MCE.
The open triangles indicate solutions of (\ref{eq:laneemden}) for 
$10^{-4}\leq a=M/4\pi \rho_{\rm m} R^3 \leq 10^{-2}$.
The boxes and asterisks indicate solutions of 
(\ref{coupled})--(\ref{coupled3}) for $\epsilon=10^{-4}$
and $10^{-3}$ respectively,
with a tolerance of $\delta=10^{-8}$.
(We have verified that the results are unchanged for
$10^{-6} \leq \delta\leq 10^{-8}$.)
Applying the procedure in the previous paragraph to compute $\rho_{\rm m}$,
we find $\log a = -10.1$ and $-9.00$ for the boxes and asterisks respectively. 
For $\xi > \xi_c$, the two models are in accord, as expected;
the halo is extended, 
so the softening (and the precise value of $a$) are not important
in the dilute regime $\rho \ll \rho_{\rm m}$.
For $\xi < \xi_c$, the two models differ appreciably.
The density of states calculation
predicts less variation of $T$ with $E$
than the van der Waals model,
for a given value of $a$.
We confirm the trend, apparent in Figure 4.11 of \citet{pad90},
that $T$ increases with $|a|$ when $\xi < \xi_c$ is fixed.
However, the trend is confirmed in the range $-10\lesssim \log a \lesssim -9$,
which does not overlap with the range $-4 \leq \log a \leq -2$
investigated by \citet{pad90}.
The effective value of $\rho_{\rm m}$
predicted by the density of states calculation is systematically
greater than anticipated in van der Waals models
published previously.

\subsection{Phase diagram on $\xi-\epsilon$ plane}

\begin{figure}
\plottwo{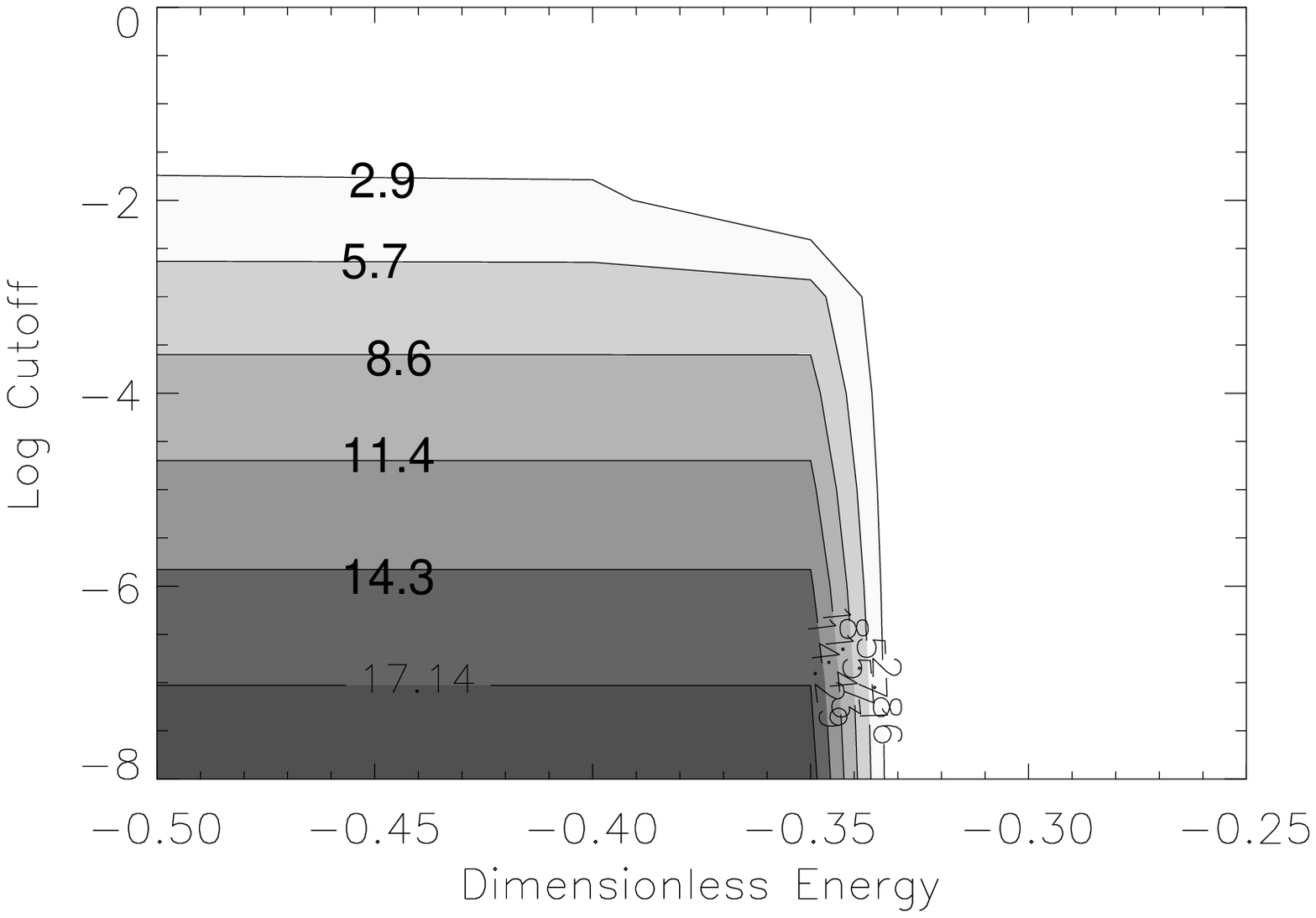}{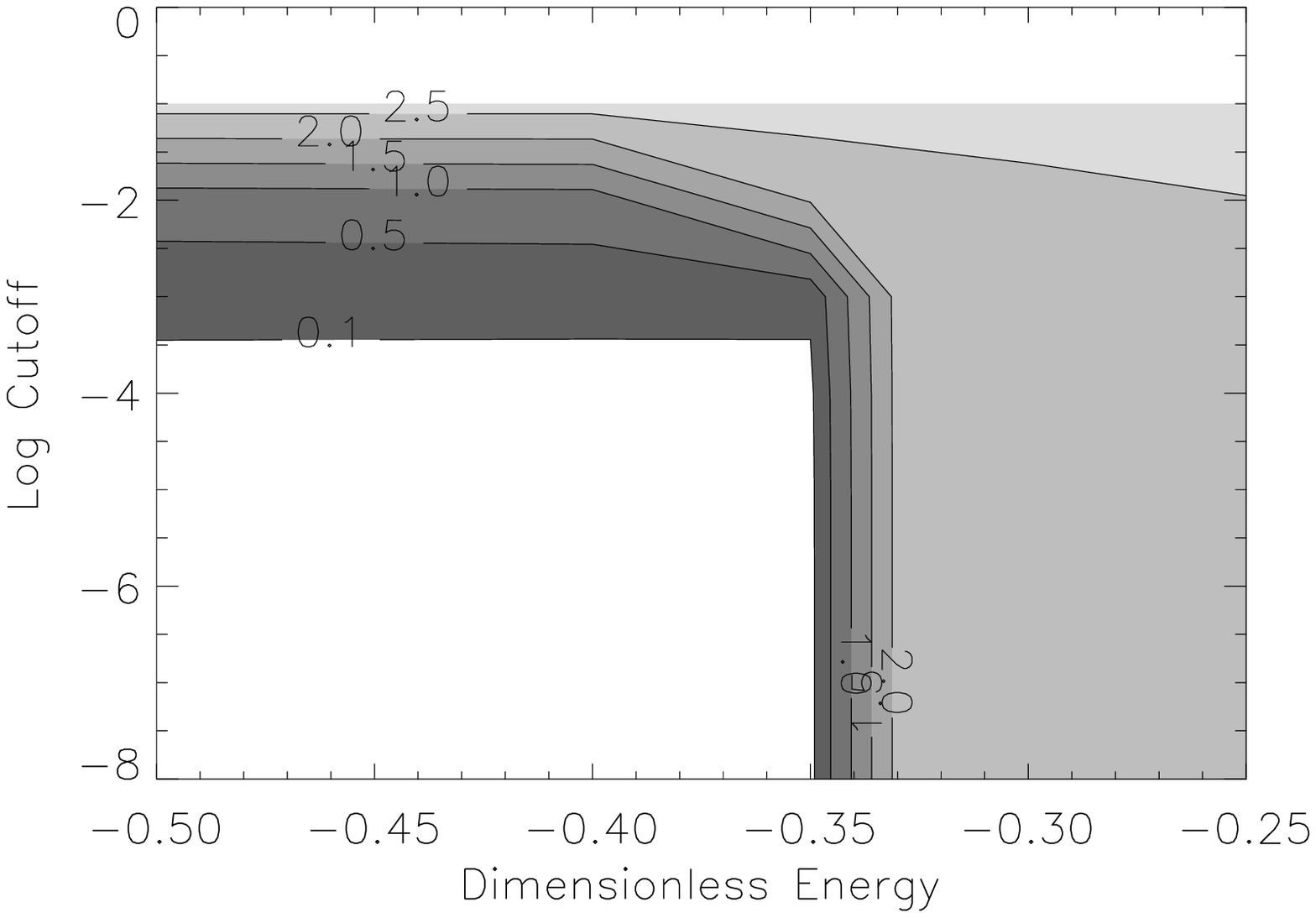}
\caption{Contour plot of the logarithmic slope , $|p|$, of the density profile in its (a) inner, $x=\epsilon$, and (b) outer, $x=1$, regions, as a function of energy, $\xi$. The phase transition is clear at $\xi\sim-0.335$ for $\epsilon\lesssim{10^{-2}}$. The inner slope $|p(\epsilon)|$ steepens as $\epsilon$ is reduced. For $\epsilon\gtrsim{10^{-2}}$, the system does not collapse. \label{contour_slope}}
\end{figure}

We can produce a phase diagram for the system by plotting the power-law exponent of the density profile as a function of $\xi$ and $\epsilon$. The logarithmic slope of the density profile is defined as

\begin{equation}
p(x) = \frac{d\ln[\rho(x)]}{d\ln{x}}.
\end{equation}
We evaluate the logarithmic slope in the inner halo, at $x=\epsilon$, and also at the box edge, $x=1$, to consistently characterise the two phases. Figures \ref{contour_slope}(a) and \ref{contour_slope}(b) display the phase diagrams obtained from $p(\epsilon)$ and $p(1)$ for $\sigma=0.01$; note that varying the value of $\sigma$ does not affect the results noticeably. The phase transition occurs where the contours are bunched, at $\xi\simeq-0.335$. For $\epsilon\lesssim10^{-2}$, $p(\epsilon)$ is large (Dirac cusp) and $p(1)$ is almost zero (flat envelope) in the collapsed phase. For $\epsilon\gtrsim{10^{-2}}$, the system does not undergo a phase transition. The entropy and temperature are continuous for all $\xi$ and the profile is extended everywhere.

%\clearpage

%\clearpage

\begin{figure}
\plotone{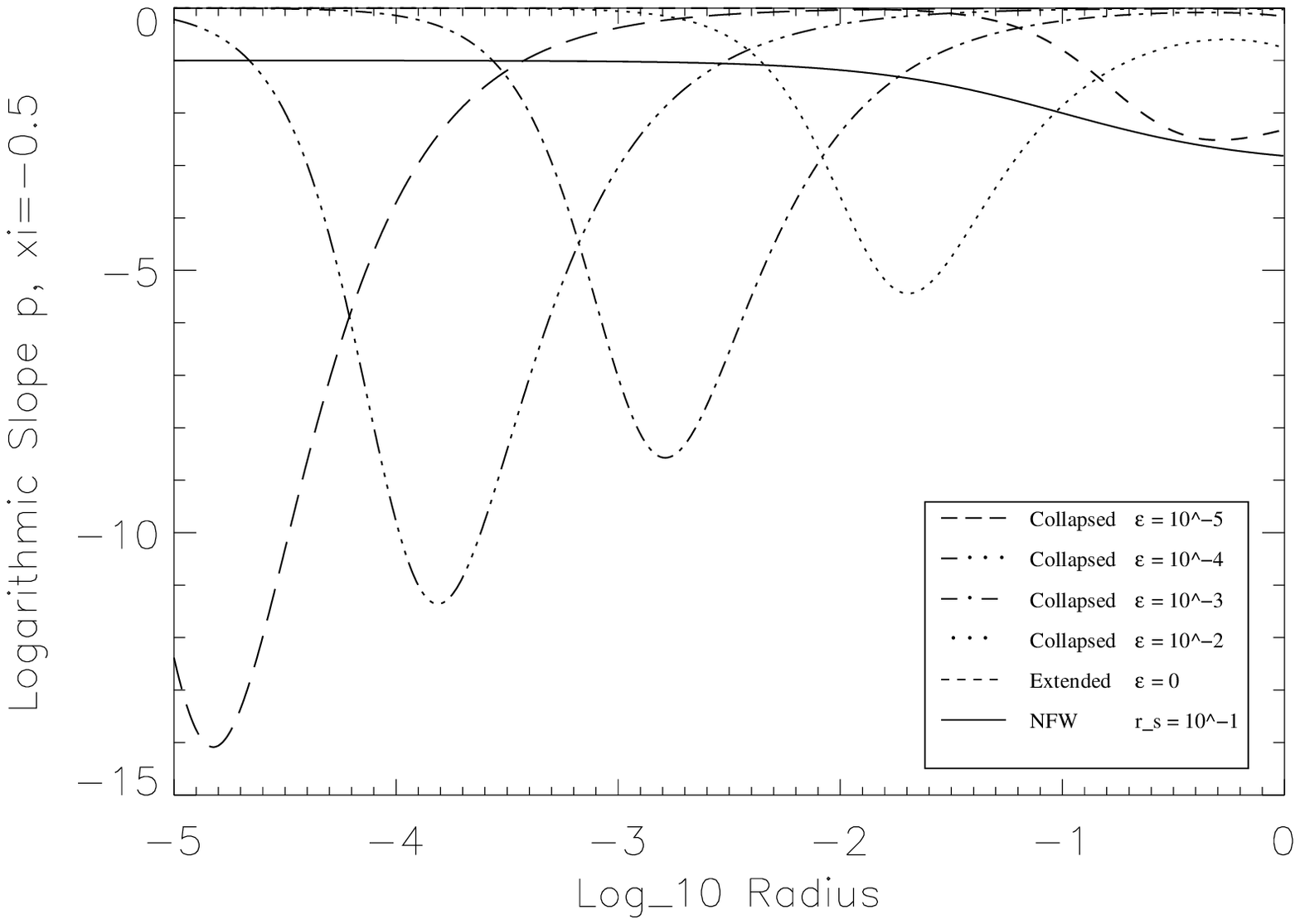}
\caption{Logarithmic slope, $p$, as a function of logarithmic radius for $\xi=-0.5$ and $\epsilon=10^{-5}$ (long dash), $10^{-4}$ (dash-dot-dot-dot), $10^{-3}$ (dash-dot) and $10^{-2}$ (dotted). The slope is a maximum at $x$ slightly above $\epsilon$. Also plotted is an NFW profile (solid) and an extended phase profile (short dash, $\xi=-0.3$). \label{log_slope}}
\end{figure}

Interestingly, the central Dirac peak steepens as $\epsilon$ decreases, in the collapsed phase. Figure \ref{log_slope} plots $p(x)$ as a function of logarithmic radius for $10^{-5}\leq\epsilon\leq10^{-2}$ and at a fixed energy $\xi=-0.5$.
The location and amplitude of the maximum of $p(x)$ is a strong function of $\epsilon$, with $p(x)\rightarrow{0}$ as $x\rightarrow{0, 1}$ for collapsed haloes. The dynamic range of numerical studies generally extends from a few times $\epsilon$ to the virial radius. In this range, $p(x)$ is strongly affected by the softening length used. The classical NFW profile \citep{nfw96} and the profile of the extended phase at $\xi>\xi_c$ are also plotted for comparison. Clearly the NFW profile, although it is cuspy, does not match exactly the collapsed (or indeed the extended) phase profile.
We are unable to explain unambiguously why the NFW profile is not reproduced, 
but remind the reader that our analysis neglects several effects, such as hierarchical clustering, 
non-zero angular momentum, and cosmological expansion,
which are present in full $N$-body simulations.
The latter effect is discussed in more detail below.

%\clearpage

%\clearpage

\subsection{Cosmological expansion}

Our calculations are performed in a static background spacetime. In the central parts of the halo, 
where we focus most of our attention, CDM particles are tightly bound and effectively decoupled 
from the cosmological expansion. Consequently, the phase transition at $\xi_c$ is essentially unaffected,
as long as the central volume where the cosmological expansion can be neglected is still large enough to
encompass the `core-halo' structure (centre-to-edge density contrast $\gtrsim 709$)
required for the gravothermal catastrophe to occur \citep{lynden68}.
From this perspective, a static background metric is a good approximation.

In the outer halo, CDM particles are loosely bound and $\rho(x)$ may be determined partly
by the expansion and certain specific cosmological parameters. 
For example, there exists a mapping between the linear and evolved two-point correlation functions of haloes 
which implies power-law density profiles with exponent $-3(n+4)/(n+5)$, 
where $n$ is the index of the power spectrum of initial fluctuations \citep{hamilton91,pad98}.
By the same token, simulations suggest that the dependence of the shape of the profile
on initial cosmological parameters is weak; this empirical result holds
for SCDM ($\Omega_M=1.0$), LCDM ($\Omega_M=0.3,\Omega_\Lambda=0.7$) and open ($\Omega_M=0.3$) cosmogonies,
as well as a wide mixture of hot and cold dark matter components \citep{huss99a}.
Note that the origin of the NFW scale radius observed in simulated haloes, $r_s$, 
is not explained by the (self-similar) \citet{pad98} mapping. However, the total power spectrum, $P(k)$, is proportional to the product of the Fourier transformed halo profile and the power spectrum of the distribution of halo centres, $P_{\rm cent}(k)$ \citep{pad02}, implying $\rho(x)\propto{x}^{-1}$ at large $k$ [deep minima of the gravitational potential, $V$, with $P_{\rm cent}(k)\propto{P_V(k)}$], and $\rho(x)\propto{x}^{-3}$ at small $k$ [quasi-linear regime, $P_{\rm cent}(k)\propto{P(k)}$] \citep{pad02}.

\citet{huss99b} demonstrated empirically that cosmological expansion does influence $\rho(x)$,
but that it is one of several relevant factors.
They simulated a halo with all tangential components of the gravitational force artificially
set to zero, reproducing the spherical infall solution with a power-law exponent of $\sim -2.2$ 
and no break in slope \citep{bert85}. They also showed that $r_s$ is not determined solely by expansion,
found evidence for the importance of angular momentum in the system,
and concluded that the unbroken power-law profiles of \citet{vanalb82} in the absence of
expansion were inadequately resolved for the purpose of testing the existence of $r_s$.
Of course, the scale radius (and hence the concentration parameter) of a halo 
does depend critically on its formation epoch:
$r_s$ is fixed by the overdensity, $\delta_c$, which is, in turn, a function of the collapse redshift
\citep{nfw97}.

\section{Comparison with numerical simulations of CDM haloes}

The key result from Section \ref{profile} from the perspective of $N$-body simulations is that any simulation with $\epsilon\neq{0}$ can produce stable haloes at energies $\xi<\xi_c$ that do not yield stable haloes in true ($\epsilon=0$) gravity. Furthermore, these stable $\epsilon\neq{0}$ haloes are collapsed, whereas stable $\epsilon=0$ haloes are extended. In this section, we show that many published $N$-body results inadvertently sample the collapsed phase only. 
We confine ourselves to studies that report the specific values of $\xi$ and $\epsilon$ explored.

\subsection{Softening length}

Simulations have been performed over a range of $N$ and with a range of resolutions in an attempt to place bounds on the optimum softening length \citep{ghigna00,splinter98,moore98}. \citet{vankampen00} argued that the choice $\epsilon \approx 0.5r_{1/2}N^{-1/3}$, where $r_{1/2}$ is the half-mass radius of the system, strikes a balance between too short a relaxation time ($\epsilon$ too small) and excessive particle clustering ($\epsilon$ too large). Similarly, \citet{athan00} find an optimal length $\epsilon=0.32N^{-0.27}$ for a $\gamma=0$ Dehnen sphere. These different criteria define a range of $\epsilon$ within which most modern simulations are performed. The softening length in high resolution simulations is 1--5 kpc for galaxy-sized haloes \citep{reed03}, corresponding to $\epsilon\sim10^{-4}$ if $R$ is taken to be the virial radius. The studies investigated below occupy the range $10^{-4}<\epsilon<10^{-2}$.

\subsection{Total energy of a halo}\label{energy}

The total energy of a simulated halo is rarely quoted in published studies. We therefore calculate $\xi$ from the quoted mass, size and concentration parameter, $c={r_s}/r_{200}$, where $r_s$ is the characteristic radius of the halo and $r_{200}=R$ is the radius at which the halo density has dropped to $200$ times the background. For a classical NFW halo, the kinetic energy, $K$, is given by \citep{mo98}

\begin{eqnarray}\label{mo}
K_{NFW} &=& \frac{GM^2}{2R}f(c),\\ 
f(c) &=& \frac{c[ 1-1/(1+c)^2-2\ln(1+c)/(1+c)]}{2 [c/(1+c)-\ln{(1+c)}]^2},
\end{eqnarray}
assuming circular orbits. The potential energy is \citep[eq. 2P-1]{bintre87}
\begin{eqnarray}
U_{NFW} &=& -\frac{GM^2}{2R}h(c),\\
h(c) &=& \int^1_0{x^{-2}[(1+cx)^{-1}-1+\ln{(1+cx)}]^2}dx,
\end{eqnarray}
assuming spherical symmetry. Similarly, for the halo profile found by \citet{moore99}, we obtain

\begin{eqnarray}
K_M &=& \frac{GM^2}{4R}\left[1 + \frac{g(c)}{\ln^2{(1+c^{3/2})}}\right],\\
U_M &=& -\frac{GM^2}{2R}\frac{g(c)}{\ln^2{(1+c^{3/2})}}, \\ \label{moore}
g(c) &=& \int^1_0 x^{-2} \ln^2{(1+c^{3/2}x^{3/2})} dx. \label{fcm}
\end{eqnarray}

We estimate the total energy, $E$, from (\ref{mo})--(\ref{fcm}) in three ways: (a) $E^{(a)}=-K$, which assumes the virial theorem and circular orbits (as stated above); (b) $E^{(b)}=U/2$, which assumes the virial theorem only; and (c) $E^{(c)}=K+U$, which assumes circular orbits, but not the virial theorem. All three approaches take $\rho({\bf x})$ to be spherically symmetric.

\subsection{Position on the phase diagram}

\begin{figure}
\plotone{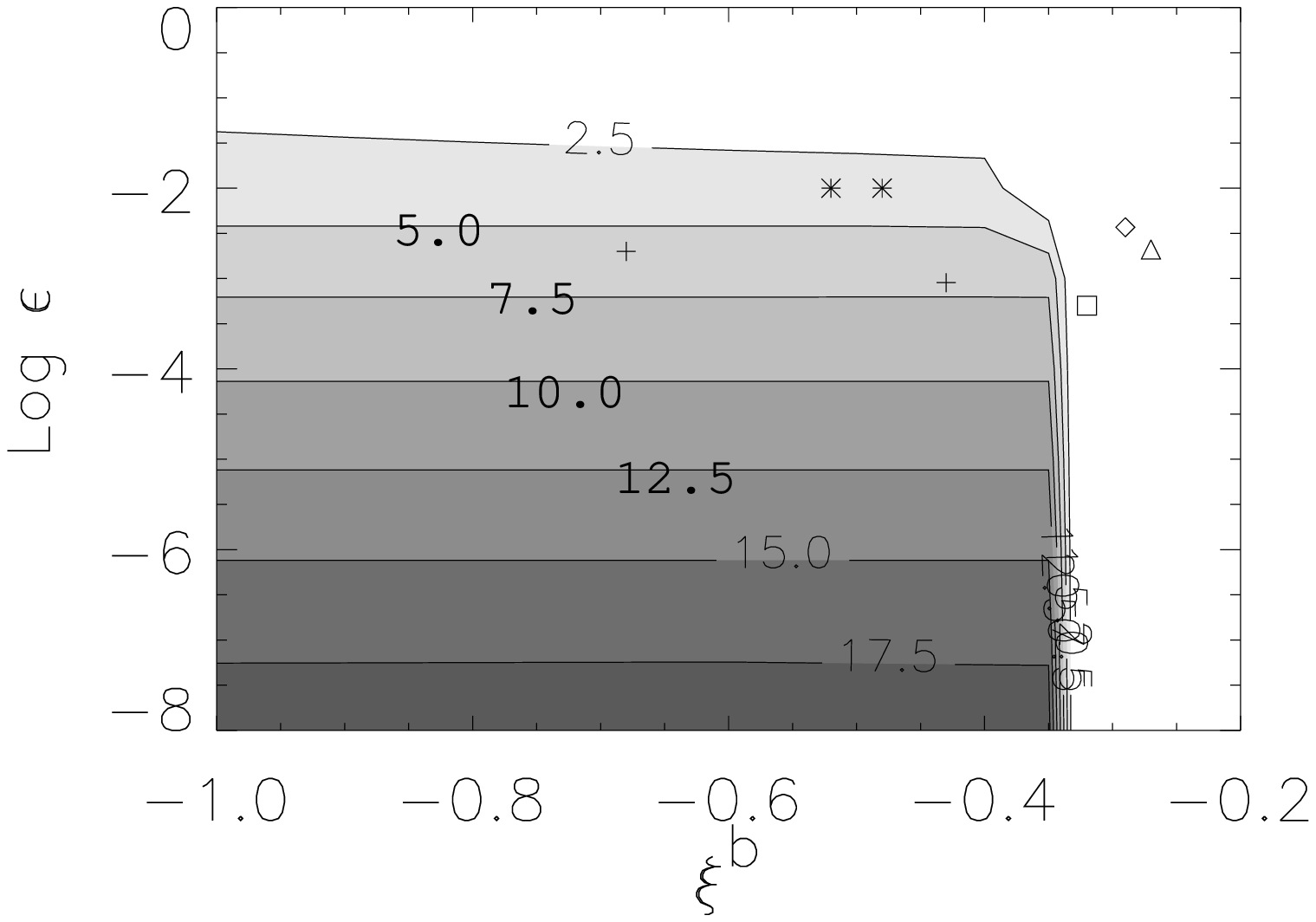}
\caption{Energy, $\xi^{(b)}$, and softening length, $\epsilon$, of published $N$-body CDM haloes overlaid on a contour plot of $p(\epsilon)$, as summarised in Table \ref{cdm_table}. The asterisks, plus signs, triangle, box and diamond denote the results of \citet{nfw96}, \citet{reed03}, \citet{hay03}, \citet{moore99} and \citet{huss99b} respectively. \label{contour_people}}
\end{figure}

We now locate on the phase diagram some examples of published $N$-body simulations of dark matter haloes in a $\Lambda$CDM cosmology, for a range of halo masses from dwarf galaxy ($\sim10^{10}M_\odot$) to cluster ($\sim10^{14}M_\odot$) size. From the output parameters, including $c$, we calculate $\xi$ and $\epsilon$ (normalised by $r_{200}$). Table \ref{cdm_table} summarises these data and the energy estimates obtained by the three methods (a)--(c) in Section \ref{energy}.
In the final column, comments are made identifying the specific haloes chosen from the referenced work.
In Figure \ref{contour_people}, we place the simulated haloes on the $\xi^{(b)}$--$\epsilon$
phase diagram [Figure \ref{contour_slope}(a)].
$\xi^{(b)}$ is preferred as it does not rely on the restrictive assumption of circular orbits. 

It is striking that the published haloes exist exclusively in the collapsed phase, 
near the left-hand edge of the diagram, or else in the bistable region,
where stable solutions exist for both collapsed and extended phases.
They do not exist in the extended phase at the right-hand edge of the diagram.
Any NFW halo with energy $\xi^{(a)}$ necessarily lies in the collapsed regime,
because one has $f(c)\geq 2/3$.
Alternatively, an NFW halo with energy $\xi^{(b)}$ lies in the collapsed regime
unless one has $c\lesssim 8$ (corresponding to more massive systems). 
Similarly, all Moore haloes have energy $\xi^{(a)}<-3/8$, and are restricted
to the collapsed regime for all $c>0$, while a Moore halo with energy $\xi^{(b)}$ is collapsed
unless one has $c\lesssim 4$. In other words, if the total energy is estimated from the
kinetic energy and virial theorem, both the NFW and Moore profiles are in the collapsed phase.
Otherwise, small concentration parameters allow either phase (in the bistable region), 
depending on the detailed route to equilibrium. 
Without the virial assumption, haloes with energy $\xi^{(c)}$ lie in the collapsed or bistable regions
as well, mostly in the former.

%\clearpage

\begin{table}
\centering
\begin{tabular}{cccccccccc}\hline
{Study}&{$M$ ($10^{12}M_\odot$)}&{$r_{200}$ (kpc)}&$c$&$\xi^{(a)}$&$\xi^{(b)}$&$\xi^{(c)}$&$\epsilon$&{Comment}&{Ref.}\\\hline
{NFW} & {2.9} & {172} & {17.54} & {-0.76} & {-0.52} & {-0.28} & {0.01} & {} & {1}\\
{} & {22.7} & {733} & {15.38} & {-0.73} & {-0.48} & {-0.23} & {0.01} & {} & {1}\\\hline
{Huss} & {500} & {1360} & {6.3} & {-0.55} & {-0.29} & {-0.03} & {0.0037} & {CDM run} & {2}\\\hline
{Moore} & {430} & {1950} & {4} & {-0.57} & {-0.32} & {-0.07} & {0.0005} & {cluster} & {3}\\\hline
{Reed} & {0.188} & {119} & {28} & {-0.93} & {-0.68} & {-0.43} & {0.0020} & {dwf1} & {4}\\
{} & {40} & {705} & {12.5} & {-0.68} & {-0.43} & {-0.18} & {0.0009} & {grp1} & {4}\\\hline
{Hayashi} & {2.2} & {212.7} & {5.3} & {-0.52} & {-0.27} & {-0.02} & {0.0021} & {G3/256$^3$} & {5}\\\hline
\end{tabular}
\caption{Recent $N$-body simulations of CDM haloes. The mass, $M$, size, $r_{200}$, concentration parameter, $c$, and softening length, $\epsilon$, are measured from the output. Equations (\ref{mo})--(\ref{fcm}) are used to calculate the halo energy in three ways, $\xi^{(a)}$, $\xi^{(b)}$ and $\xi^{(c)}$. All calculations assume an NFW profile, except for \citet{moore99} which uses the Moore profile. \label{cdm_table}}
\tablerefs{(1) \citet{nfw96}; (2) \citet{huss99b}; (3) \citet{moore99}; (4) \citet{reed03}; (5) \citet{hay03}}
\end{table}

%\clearpage

%\clearpage

\section{Conclusion}

We have investigated the equilibrium configurations of $N$ self-gravitating collisionless particles,
interacting via a softened gravitational potential, in the MCE and mean-field limit. 
Below a critical energy, $\xi_c \simeq -0.335$, a system with $\epsilon\neq 0$ exists in a stable, collapsed phase.
This phase is unstable for pure gravity ($\epsilon=0$). 
Above another critical energy $\xi_c^{(+)} \simeq 0$,
both softened and unsoftened systems exist in an stable, extended phase. 
In the intermediate region $\xi_c < \xi < \xi_c^{(+)}$, both the collapsed and extended phases
are accessible; the detailed route to equilibrium determines which one is picked out.
The density profiles for the extended and collapsed phases are qualitatively different. 
The extended profile has a flat core and near-isothermal outer envelope. 
The collapsed profile is a centrally condensed Dirac peak, whose logarithmic slope 
depends on $\epsilon$.

We compare our results with published $N$-body simulations by using the softening parameter, 
$\epsilon$, size, $r_{200}$, and concentration parameter, $c$, to place simulated haloes on the 
$\xi$-$\epsilon$ phase diagram. We find that many published simulations inadvertently sample 
the collapsed phase only, even though this phase is unstable for pure gravity and
arguably irrelevant astrophysically.

We remind the reader that we neglect several effects that are important in real CDM haloes, such as hierarchical clustering, nonzero angular momentum, and cosmological expansion. Our results elucidate some of the artificial behaviour that a softened potential can introduce; they are not a substitute for a full $N$-body calculation.

\acknowledgments
We are grateful to Bruce McKellar for extensive discussions on the theoretical basis of the mean-field equations, and the anonymous referee for useful comments that improved the manuscript. This work was supported by the Australian Research Council Discovery Project grant 0208618. CMT acknowledges the funding provided by an Australian Postgraduate Award.

%\clearpage

\end{document}